\documentclass[conference]{IEEEtran}
\IEEEoverridecommandlockouts

\usepackage[style=ieee,backend=bibtex]{biblatex}
\DeclareFieldFormat{doi}{}                
\usepackage{amsmath,amssymb,amsfonts}
\usepackage{graphicx}
\usepackage{textcomp}
\usepackage{xcolor}

\usepackage{algorithm}
\usepackage{algorithmic}

\usepackage[caption=false,font=footnotesize]{subfig}

\usepackage{cleveref}
\crefname{figure}{Fig.}{Figs.}
\Crefname{figure}{Fig.}{Figs.}
\crefname{equation}{Eq.}{Eqs.}
\Crefname{equation}{Eq.}{Eqs.}
\crefname{appendix}{Appendix}{Appendices}
\Crefname{appendix}{Appendix}{Appendices}
\crefname{section}{Section}{Sections}
\Crefname{section}{Section}{Sections}
\begin{document}

\title{Efficient Near Field Beam Tracking via Thompson Sampling}

\author{
Junchi Liu\textsuperscript{\dag},
Zijun Wang\textsuperscript{\dag},
Shawn Tsai\textsuperscript{\ddag},
Rui Zhang\textsuperscript{\dag}\\[0.5ex]
\textsuperscript{\dag}Department of Electrical Engineering, University at Buffalo, SUNY, Buffalo, NY, USA\\
Emails: junchili@buffalo.edu, zwang267@buffalo.edu, rzhang45@buffalo.edu\\
\textsuperscript{\ddag}CSD, MediaTek Inc. USA, San Diego, CA, USA. Email: shawn.tsai@mediatek.com
}
\maketitle

\begin{abstract}
The shift to the radiative near-field region due to large antenna arrays necessitates beamforming that accounts for both angle and range, evolving mobility management into a joint angular–range tracking challenge. Conventional schemes rely on rigid pilot–payload structures with dedicated training slots, which interrupt data transmission and degrade spectral efficiency. To address this, we propose a pilot-free beam tracking framework leveraging Thompson sampling(TS).
Within each sliding window, the user trajectory is modeled by local low-order polynomials in angle and range, and the motion parameters are estimated by maximum likelihood with uncertainty quantified via the Fisher information matrix. 
TS adaptively probes uncertain trajectory regions using beams that simultaneously serve as payload beams. Simulations demonstrate that the proposed framework maintains reliable connectivity while eliminating the overhead of dedicated pilot-based beam sweeping.
\end{abstract}


\section{Introduction}
\label{sec:intro}
\noindent Extremely large aperture arrays (ELAAs) and higher frequency bands, such as mmWave and THz, are key enablers of 6G networks. As the array aperture expands, the radiative near-field region grows significantly, often covering the typical service area of user equipments (UEs) \cite{Tutorial_with_beam_pattern_and_resolution}.
In this near-field regime, the channel steering vector becomes dependent on both angle and range, contrasting with the far-field assumption where only the angular coordinate is relevant \cite{far-field-assumption,cui2022farornf}.
Furthermore, user mobility in the radiative near-field transforms beam management into a coupled angular–range tracking problem. While far-field beam tracking is well-studied using model-based estimators like the Extended Kalman Filter (EKF) \cite{EKF2016ICC} or data-driven deep learning approaches such as Long Short-Term Memory (LSTM) and Liquid Neural Networks (LNNs) \cite{LNN_tracking}, the pilot overhead will be particularly significant in near field due to the increased dimensionality of the angle-range search space \cite{cui2022farornf}.  
There have been efforts toward efficient and low-overhead near-field beam tracking recently. For instance, \cite{UPA_NF} estimates the effective beam coherence time and builds a dynamic non-uniform coordinate searching grid based on the prior predictions. Therefore, such a system enables refined local tracking rather than an exhaustive search over the whole codebook. 
\cite{sub_array_li} leverages sub-array architectures combined with EKF to predict the user's position and velocity with one uplink pilot transmission each tracking cycle.
Existing tracking protocols typically rely on the pilot–prediction–pilot structure. In these schemes, periodic beam sweeping or dedicated pilot transmissions interrupt payload delivery, taking away certain spectral resources from data transmission.

To improve system efficiency, we propose a pilot-free continuous tracking framework. Specifically, we model the physical motion parameters (i.e., angle and range) over a sliding observation window using piecewise-polynomial functions to capture kinematics over short time intervals. This model allows for channel reconstruction while naturally adapting to varying motion dynamics. Based on this formulation, we derive a Maximum Likelihood (ML) estimator for the motion parameters and characterize its statistical distribution via the Fisher Information Matrix (FIM). Building on this characterization, we develop a TS-based tracking strategy that adaptively selects the next beam probe by sampling from the posterior distribution over the motion parameters. This approach enables the system to balance exploration and exploitation, effectively using data beams as probes to maintain reliable connectivity without the need for dedicated pilot overhead.
The paper is organized as follows. \Cref{sec:sysprob} describes the system model and formulate the MLE problem. \Cref{sec:method} introduces our proposed TS-based tracking algorithm, which dynamically balances exploration and exploitation. \Cref{sim} presents the simulation setup and compares the performance of our method with benchmark schemes. Finally, \Cref{sec:con} provides concluding remarks.



\section{System Model And Problem Formulation}
\label{sec:sysprob}
This section introduces the near-field channel model for a mobile UE and formulates the beam tracking problem under such channel conditions.

\subsection{Channel Model}\label{subsec:channel}
\begin{figure}[t]
\centering
\includegraphics[width=\linewidth]{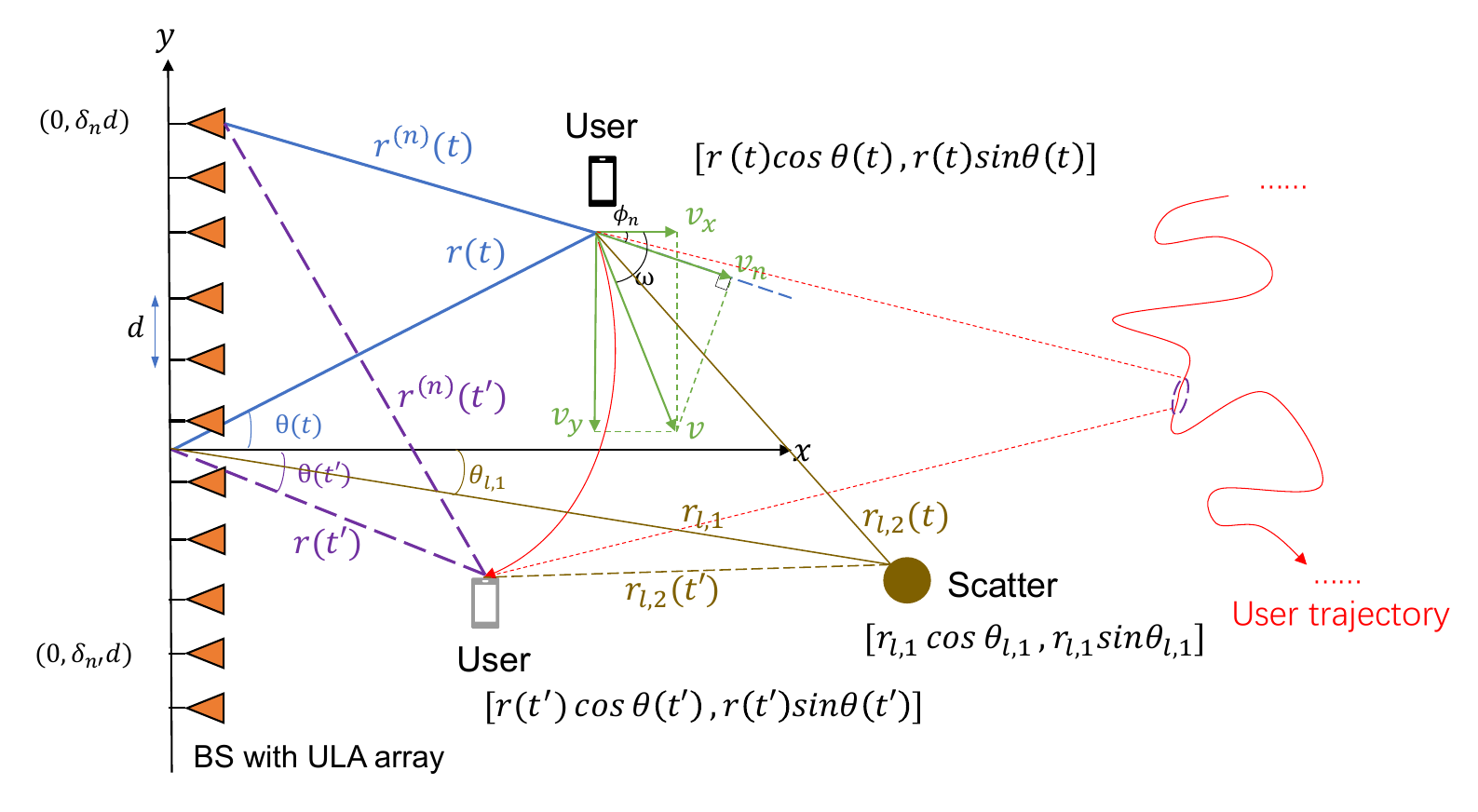}
\caption{Diagram of ULA based near field channel with mobile user and scatterer.}
\label{channel}
\end{figure}

We consider a near-field multiple-input and single-output (MISO) downlink system as illustrated in Fig.~\ref{channel}, where the base station (BS) is equipped with an $N$-elements uniform linear array (ULA) and the moving UE has a single antenna. 
Following the criterion in \cite{fresnel}, we characterize the near-field region as the distance interval between the Fresnel distance and the Rayleigh distance, i.e.,
$R_{\mathrm{Fre}}=\frac{1}{2}\sqrt{\frac{D^{3}}{\lambda}}$ and $R_{\mathrm{Ray}}=\frac{2D^{2}}{\lambda}$,
respectively. Here, $\lambda$ denotes the carrier wavelength, and $D=(N-1)d$ represents the physical aperture of the ULA with inter-element spacing $d$.
At time $t$, the BS transmits a symbol $x_t$ using a beamformer $\mathbf{w}_t\in\mathbb{C}^{N\times 1}$, and the UE observes
\begin{equation}\label{eq:y_model}
y_t = \mathbf{h}_{t}^{\mathsf H}\mathbf{w}_t x_t + n_t,
\end{equation}
where $n_t\sim\mathcal{CN}(0,\sigma^2)$ and $\mathbf{h}_{t}\in\mathbb{C}^{N}$ is the channel vector at $t$. $x_t$ denotes the symbol transmitted at time $t$.

As depicted in Fig.~\ref{channel}, the BS employs a ULA aligned with the $y$-axis and centered at the origin $(0,0)$. The position of the $n$-th antenna element is given by $(0,\delta_n d)$, where 
$\delta_n \triangleq \frac{2n-N+1}{2}$, $n\in\mathcal{N}$, and $\mathcal{N}\triangleq\{0,1,\ldots,N-1\}$. 
The inter-element spacing is set to $d=\lambda/2$. 
The line-of-sight (LoS) channel vector $\mathbf{h}^{LoS}_{t}$ is determined by the user's time-varying position, which is represented by the polar coordinates $[\theta(t), r(t)]$. Specifically, $r(t)$ and $\theta(t)$ denote the instantaneous range and angle of the user relative to the center of the BS array, respectively. To express this explicitly, we first define the near-field array steering vector $\mathbf{b} (\theta(t),r(t))$ associated with UE:

\begin{IEEEeqnarray}{rCl}
\mathbf{b}\!\big(\theta(t),r(t)\big)
&=& \frac{1}{\sqrt{N}}
\Big[
e^{-j\frac{2\pi}{\lambda}\big(r^{(0)}(t)-r(t)\big)},\nonumber\\
&&\; \ldots,\ 
e^{-j\frac{2\pi}{\lambda}\big(r^{(N-1)}(t)-r(t)\big)}
\Big]^{\mathsf T}.
\end{IEEEeqnarray}
where the propagation distance $r^{(n)}(t)$ from the $n$-th antenna element to the UE is defined as:  
\begin{equation}\label{eq:rn}
r^{(n)}(t)=\sqrt{r^2(t)+\delta_n^2 d^2-2r(t)\sin(\theta(t))\delta_n d},
\end{equation}


Accordingly, the BS--UE LoS channel vector at $t$ can be modeled as 
\begin{equation}\label{eq:h_main}
\begin{split}
\mathbf{h}^{LoS}_{t} &= g(t)\,e^{-j\frac{2\pi r(t)}{\lambda}}\cdot\ \mathbf{b}(\theta(t),r(t)),
\end{split}
\end{equation}
where $g(t) \triangleq \frac{\lambda}{4\pi r(t)}$ captures the large-scale path gain due to free-space path loss \cite{sun2016pathloss}. 
We assume that scatterers in the environment remain stationary during the tracking process. Consequently, the non-line-of-sight (NLoS) channel component at time $t$ is modeled as a sum over the $L-1$ static reflection paths \cite{LOS_NLos_dominated_Tutorial}:
\begin{equation}
    \mathbf{h}^{NLoS}_{t} = \sum_{l=1}^{L-1} g_l e^{-j \frac{2\pi(r_{l,1}+r_{l,2})}{\lambda}} \mathbf{b}(\theta_l, r_{l,1}),
    \label{eq:nlos}
\end{equation}
where $L$  is the total number of propagation paths,  the $l$-th scatter is located at polar coordinates $(\theta_l, r_{l,1})$ relative to the center of BS antenna array. From the geometry shown in Fig.~\ref{channel}, the distance from the scatterer to the UE is $r_{l,2} = \sqrt{r(t)^2 + r_{l,1}^2 - 2r(t) r_{l,1} \cos(\theta(t) - \theta_l)}$. $g_l = \frac{\lambda p_l}{4\pi (r_{l,1} \cdot r_{l,2})}$ is the NLoS channel gain that includes the path loss and the reflection character of $l$-th scatter $p_l$. Then, the complete multi-path channel at time $t$ is given by the sum of the LoS and NLoS components 
 \begin{equation}\label{eq:sum_channel}
     \mathbf{h}_{t} = \mathbf{h}^{LoS}_{t } + \mathbf{h}^{NLoS}_{t }.
 \end{equation}
\subsection{Problem Formulation}\label{subsec:prob-form}

Given the dominance of LoS propagation in high-frequency bands \cite{NYU_NLOS}, we tracks only the LoS path in the following analysis without loss of generality, i.e. $\hat{\mathbf{h}}_{t } = \mathbf{\hat{h}}^{LoS}_{t }$.
The objective is to track the time-varying polar coordinates of the UE, i.e., $\big(\theta(t),r(t)\big)$, and accordingly update the transmit beamformer $\mathbf{{w}}_{t}$ to sustain beam alignment. 


\begin{figure}[!t]
\centering
\includegraphics[width=\linewidth]{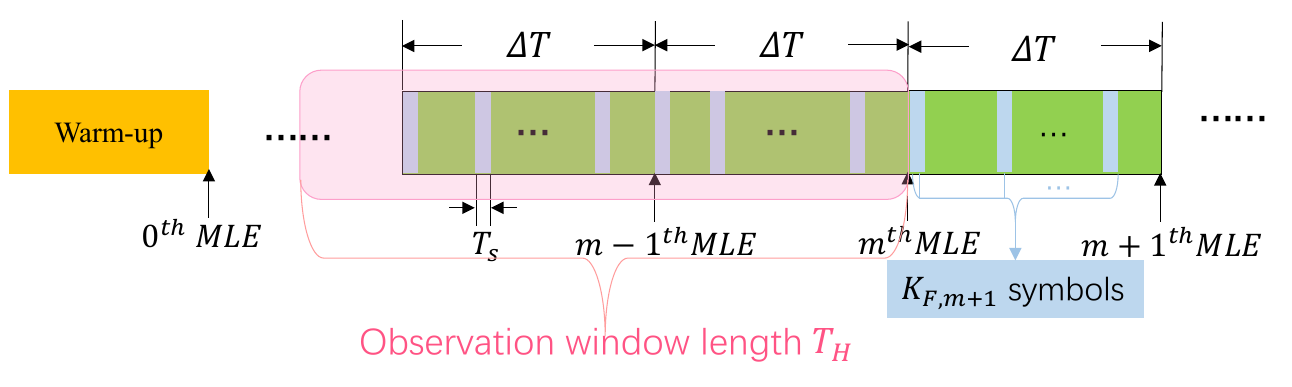} 
\caption{Proposed beam tracking protocol.}
\label{fig:tracking_protocol}
\end{figure}

Assuming that the UE’s trajectory is locally smooth over a time interval (i.e., the observation window for MLE), whose length is $T_H$. This ensures that the polynomial model remains an accurate representation of the UE's instantaneous kinematics throughout the estimation process. We approximate the temporal evolution of the angle and range using low-order polynomials in every sliding window. Thus, at local time $t'$ ( $t'=0$ denotes the local time index at the beginning of the sliding window),  the instantaneous state $\left [ \hat{\theta}(t') ,\hat{r}(t') \right ]$ can be expressed as 
\begin{equation}\label{eq:poly_apprx}
\hat{\theta}(t') = \sum_{i=0}^{p_\alpha} {\alpha}_i t'^i, \quad
\;
\hat{r}(t')= \sum_{i=0}^{p_\beta}  {\beta}_i t'^i,
\end{equation}
where ${\boldsymbol{\alpha}}\triangleq[{\alpha}_0,\ldots,{\alpha}_{p_\alpha}]^{\mathsf T}$ and
${\boldsymbol{\beta}}\triangleq[{\beta}_0,\ldots,{\beta}_{p_\beta}]^{\mathsf T}$ represents the unknown polynomial coefficients that need to be estimated by MLE. Hence, the steering vector in Eq. \eqref{eq:h_main} at time $t'$ can be rewritten as: 
\begin{equation}\label{eq:h_est}
\hat{\mathbf{h}}^{LoS}_{t'} = \hat{g}(t')\,e^{-j\frac{2\pi \hat{r}(t')}{\lambda}}\cdot\ \mathbf{b}(\hat{\theta}(t'),\hat{r}(t')).
\end{equation}

Following an initial warm-up phase, during which sufficient observations are collected to initialize the first MLE update, the system enters the tracking protocol illustrated in \Cref{fig:tracking_protocol}. The tracking process is discretized into a sequence of update cycles. 
The time interval between two consecutive MLEs spans a duration $\Delta T$ and contains $K=\Delta T/T_s$ transmitted payload symbols, where $T_s$ denotes the symbol duration. Within each MLE interval, the UE feeds back $K_F \leq K$ received symbols to the BS for the MLE update. To ensure that the feedback captures the temporal evolution of the channel and prevents trajectory aliasing, these $K_F$ symbols are selected at predefined positions that are approximately uniformly distributed over the $K$ transmitted symbols. Specifically, we define the feedback-position set in one MLE interval as:

\begin{equation}
    \mathcal{K}_F \triangleq \left\{\, 1+\left\lfloor \frac{(i-1)(K-1)}{K_F-1} \right\rfloor \;\middle|\; i=1,2,\ldots,K_F \right\}.
\end{equation}
After the warm-up stage, we denote $m$ as the MLE interval index, and $k$ as the inner symbol index in each interval, where $k \in \{1,2,\ldots,K\}$. The global symbol index is defined as $u \triangleq u(m,k)=(m-1)K+k$, and its corresponding time is denoted as $t'_u \triangleq uT_s$. When $k \in \mathcal{K}_F$, the $u(m,k)$-th symbol is fed back to the BS.

At the end of $m$-th MLE interval, let $\mathcal{P}_m$ denote the set of feedback symbol indices from $m\Delta T-T_H$ to $m\Delta T$, the history dataset available for $m$-th MLE is defined as:
\begin{equation}\label{eq:history_def}
\mathcal{H}_m \triangleq \big\{ (y_{u}, \mathbf{w}_{u}) \big\}_{u \in \mathcal{P}_m},
\end{equation}
where $y_{u}$ is the received feedback signal and $\mathbf{w}_{u}$ is the beamformer transmitted at slot $u$. For any symbol index $u \in \mathcal{P}_m$, at $t' = t'_{u}$, the estimated channel in Eq. \eqref{eq:h_est} is denoted as $\mathbf{\hat{h}}_{u }$ correspondingly. Then the estimated noiseless observation model is:
\begin{equation}\label{eq:mu_def}
\mu_{u}(\boldsymbol{{\alpha}},\boldsymbol{{\beta}})
\triangleq \mathbf{\hat{h}}^{\mathsf H}_{u }\mathbf{w}_{u},
\end{equation}
Assuming the observations are subject to additive white Gaussian noise (AWGN) with variance $\sigma^2$, the conditional probability density function (PDF) of the received signal is:
\begin{equation}\label{eq:cond_pdf}
p\!\left(y_{u}\mid \boldsymbol{{\alpha}},\boldsymbol{{\beta}}\right)
=\frac{1}{\pi\sigma^2}\exp\!\left(
-\frac{\left|y_{u}-\mu_{u}(\boldsymbol{{\alpha}},\boldsymbol{{\beta}})\right|^2}{\sigma^2}
\right).
\end{equation}

Assuming independent observations over the sliding window, the joint likelihood function is:

\begin{equation}\label{eq:likelihood}
\begin{split}
\mathcal{L}_m(\boldsymbol{{\alpha}},\boldsymbol{{\beta}})
&\triangleq \prod_{u\in\mathcal{P}_m} p\!\left(y_{u}\mid \boldsymbol{{\alpha}},\boldsymbol{{\beta}}\right) \\
&= (\pi\sigma^2)^{-|\mathcal{P}_m|}  \exp\!\Bigg( -\frac{1}{\sigma^2}\sum_{u\in\mathcal{P}_m} 
\left|y_{u}-\mu_{u}(\boldsymbol{{\alpha}},\boldsymbol{{\beta}})\right|^2 \Bigg).
\end{split}
\end{equation}

Maximizing the likelihood function in Eq. \eqref{eq:likelihood} is equivalent to minimizing the negative log-likelihood, resulting in the following non-linear least-squares problem:
\begin{equation}\label{eq:mle_problem}
(\hat{\boldsymbol{\alpha}}_m,\hat{\boldsymbol{\beta}}_m)
= \arg\min_{\boldsymbol{\alpha},\boldsymbol{\beta}} \;
J_m({\boldsymbol{\alpha}},{\boldsymbol{\beta}}),
\end{equation}
where
\begin{equation}\label{eq:cost_J}
J_m({\boldsymbol{\alpha}},{\boldsymbol{\beta}})
\triangleq \sum_{u\in\mathcal{P}_m} \left|y_{u}-\mu_{u}({\boldsymbol{\alpha}},{\boldsymbol{\beta}})\right|^2 .
\end{equation}


Upon completing the payload transmission of each tracking cycle, the BS also finishes estimating the UE's motion parameters by solving Eq. \eqref{eq:mle_problem}. 
Due to the non-linear relationship between the channel vector and the polynomial coefficients, the objective function $J_m(\boldsymbol{{\alpha}},\boldsymbol{{\beta}})$ is generally non-convex. Thus, we solve Eq. \eqref{eq:mle_problem} using the Adam optimizer, which leverages adaptive first and second-order moment estimates to ensure robust convergence. To accelerate algorithm convergence, the results from the previous MLE are utilized as initial values for the subsequent MLE estimation.

For $\boldsymbol{\vartheta}\in\{\boldsymbol{{\alpha}},\boldsymbol{{\beta}}\}$, let the gradient at $\ell$-th iteration be defined as
$\boldsymbol{g}_{\boldsymbol{\vartheta}}^{(\ell)}\triangleq\nabla_{\boldsymbol{\vartheta}}J_m(\boldsymbol{{\alpha}}^{(\ell)},\boldsymbol{{\beta}}^{(\ell)})$.
Given shared decay factors $(\rho_1,\rho_2)\in(0,1)^2$ and parameter-specific step-sizes $\eta_{\boldsymbol{\alpha}}=\eta_{\alpha}$ and $\eta_{\boldsymbol{\beta}}=\eta_{\beta}$, the Adam recursion at iteration $\ell$ is given by:

\begin{equation}\label{eq:adam_updates}
\begin{aligned}
\boldsymbol{m}_{\boldsymbol{\vartheta}}^{(\ell)} &= \rho_1 \boldsymbol{m}_{\boldsymbol{\vartheta}}^{(\ell-1)} + (1-\rho_1)\boldsymbol{g}_{\boldsymbol{\vartheta}}^{(\ell)},\\
\boldsymbol{v}_{\boldsymbol{\vartheta}}^{(\ell)} &= \rho_2 \boldsymbol{v}_{\boldsymbol{\vartheta}}^{(\ell-1)} + (1-\rho_2)\big(\boldsymbol{g}_{\boldsymbol{\vartheta}}^{(\ell)}\odot \boldsymbol{g}_{\boldsymbol{\vartheta}}^{(\ell)}\big),\\
\tilde{\boldsymbol{m}}_{\boldsymbol{\vartheta}}^{(\ell)} &= \boldsymbol{m}_{\boldsymbol{\vartheta}}^{(\ell)}/(1-\rho_1^\ell),\;
\tilde{\boldsymbol{v}}_{\boldsymbol{\vartheta}}^{(\ell)} = \boldsymbol{v}_{\boldsymbol{\vartheta}}^{(\ell)}/(1-\rho_2^\ell),\\
\boldsymbol{\vartheta}^{(\ell+1)} &= \boldsymbol{\vartheta}^{(\ell)} - \eta_{\boldsymbol{\vartheta}}\,
\tilde{\boldsymbol{m}}_{\boldsymbol{\vartheta}}^{(\ell)} \oslash \sqrt{\tilde{\boldsymbol{v}}_{\boldsymbol{\vartheta}}^{(\ell)}} .
\end{aligned}
\end{equation}

\noindent where $\boldsymbol{m}_{\boldsymbol{\vartheta}}^{(\ell)}$ and $\boldsymbol{v}_{\boldsymbol{\vartheta}}^{(\ell)}$ are the (biased) first- and second-moment estimates of the gradient, $\tilde{\boldsymbol{m}}_{\boldsymbol{\vartheta}}^{(\ell)}$ and $\tilde{\boldsymbol{v}}_{\boldsymbol{\vartheta}}^{(\ell)}$ are their bias-corrected versions.



\section{Methodology}
\label{sec:method}
This section introduces two beam selection strategies for the payload transmission. 

\subsection{Pure Exploitation}\label{exploit}
The pure exploitation strategy focuses exclusively on transmitting the directions predicted by the current maximum-likelihood estimates obtained from Eq. \eqref{eq:mle_problem}-Eq. \eqref{eq:adam_updates}. This approach aims to maintain high beamforming gain based on the estimation results. 
In this scheme, the BS designs the beamformer during the $(m+1)$-th transmission interval by exclusively utilizing trajectory parameters derived from the MLE at the end of $m$-th interval, i.e., $ (\hat{\boldsymbol{\alpha}}_{m},\hat{\boldsymbol{\beta}}_{m})$ in Eq. \eqref{eq:mle_problem}. In this case, 
 \begin{equation}\label{eq:pillot-2}
 \mathbf{w}_u
 =\frac{\widehat{\mathbf{h}}_{u }^{(m)}}{\left\|\widehat{\mathbf{h}}_{u }^{(m)}\right\|_2},
 \end{equation}
where $\widehat{\mathbf{h}}_{u }^{(m)}$ denotes the predicted channel vector at $t'_u$ from $m$-th MLE. While this strategy preserves connectivity and may allow data transmission for the following symbols, it yields limited new information about other potential directions. Consequently, the absence of exploration can easily lead to track loss once the predicted path diverges from the actual UE trajectory.

\subsection{Proposed TS-based beam tracking scheme}\label{sec:TS}
Thompson Sampling(TS) is a Bayesian randomized decision-making strategy that naturally balances exploration and exploitation. It operates by sampling a set of parameters from their current posterior distribution and then selecting the action that is optimal under the sampled instance~\cite{russo2018tutorial}. This approach is particularly suitable for beam tracking in near-field scenarios, as it allows adaptive data beam probing without relying on pilots by minimizing cumulative regret, thereby maintaining reliable connectivity while efficiently acquiring channel information.

To enable efficient sampling, we use a multivariate Gaussian to approximate the posterior distribution in $(m+1)$-th data transmission intervals:
\begin{equation}\label{eq:post_dist}
    p \big( (\boldsymbol{\alpha}_u, \boldsymbol{\beta}_u) \mid \mathcal{H}_m \big) \approx \mathcal{N} \big( (\hat{\boldsymbol{\alpha}}_m, \hat{\boldsymbol{\beta}}_m), \boldsymbol{\Sigma}_{\alpha, \beta; m}  \big),
\end{equation}
where $u=u(m,k)$, and $\boldsymbol{\Sigma}_{\boldsymbol{\alpha}, \boldsymbol{\beta}; m}$ is the inverse of the observed Fisher information matrices (FIMs) :
\begin{equation}\label{eq:fim}
\begin{aligned}
\boldsymbol{\Sigma}_{\alpha, \beta; m} &= \mathcal{I}_{\alpha, \beta; m}^{-1} \triangleq -\frac{1}{\sigma^2}\nabla_{\boldsymbol{\alpha}, \boldsymbol{\beta}}^{2} J_m.
\end{aligned}
\end{equation}
After sampling the $\boldsymbol{\tilde{\alpha}}_{u}$ and $\boldsymbol{\tilde{\beta}}_{u}$ from the current posterior distribution at time slot $u$, the beamformer is constructed as $\mathbf{w}_u =\widetilde{\mathbf{h}}_{u }^{(m)} / ||\widetilde{\mathbf{h}}_{u }^{(m)} ||_2$, 
where the sampled channel $\widetilde{\mathbf{h}}_{u }^{(m)}$ is obtained after substituting ${\boldsymbol{\alpha}}$, ${\boldsymbol{\beta}}$ with $\boldsymbol{\tilde{\alpha}}_{u}$, $\boldsymbol{\tilde{\beta}}_{u}$ from Eq. \eqref{eq:poly_apprx}.
These covariance matrices inherently facilitate a balance between exploration and exploitation. Parameter dimensions characterized by high uncertainty (i.e., manifesting as larger posterior variances) are sampled more extensively by the TS framework. This mechanism ensures that the system exploits the estimated trajectory when confidence is high (indicated by a small $\boldsymbol{\Sigma}_m$), while autonomously initiating exploration in alternative directions as uncertainty increases.

In the proposed protocol, symbols designated for feedback utilize probing beams derived from TS to explore the angular–range uncertainty. Conversely, symbols that do not require feedback are transmitted using beams focused on pure exploitation of the current trajectory estimate, thereby maximizing the effective communication gain. 




\section{Simulation results}\label{sim}
\subsection{Simulation Settings}

The simulation is conducted with a 256-element ULA operating at $f_c=73$ GHz. The total simulation time is set to $T=4$ s with a symbol duration of $T_s = 1/30$kHz.
To test tracking robustness against non-uniform motion, the UE moves within a confined near-field region ($r \in [8, 80]$ m, $\theta \in [-\pi/3, \pi/3]$) with a time-varying velocity, whose average speed is $3.11$ m/s and maximum speed is $4.712$ m/s. The proposed tracking scheme uses a polynomial motion model with orders $p_{\alpha}=3$ for angle and $p_{\beta}=6$ for range, utilizing a history window of $T_H = 66.6$ ms and the SNR is set to 20 dB.

In our simulation, as the relative scale of $\boldsymbol{\alpha}$ and $\boldsymbol{\beta}$ differ significantly, for numerical stability, we sample $\tilde{\boldsymbol{\alpha}}$ and $\tilde{\boldsymbol{\beta}}$ from the posterior distributions independently, ignoring their correlation. In particular, at each round, we independently sample
\begin{equation}
\boldsymbol{\tilde{\alpha}}_{u}\sim\mathcal{N}\!\left(\hat{\boldsymbol{\alpha}}_{m},\boldsymbol{\Sigma}_{\alpha,m}\right),
\;
\boldsymbol{\tilde{\beta}}_{u}\sim\mathcal{N}\!\left(\hat{\boldsymbol{\beta}}_{m},\boldsymbol{\Sigma}_{\beta,m}\right),
\end{equation}
where $\Sigma_{\alpha, m} = - \tfrac{1}{\sigma^2} \nabla_{\hat{\boldsymbol{\alpha}}_m}^{2} J_m$ and $\Sigma_{\beta, m} = - \tfrac{1}{\sigma^2} \nabla_{\hat{\boldsymbol{\beta}}_m}^{2} J_m$.

\subsection{Performance Analysis}
\begin{figure}[t]
    \centering
    \includegraphics[width=\linewidth]{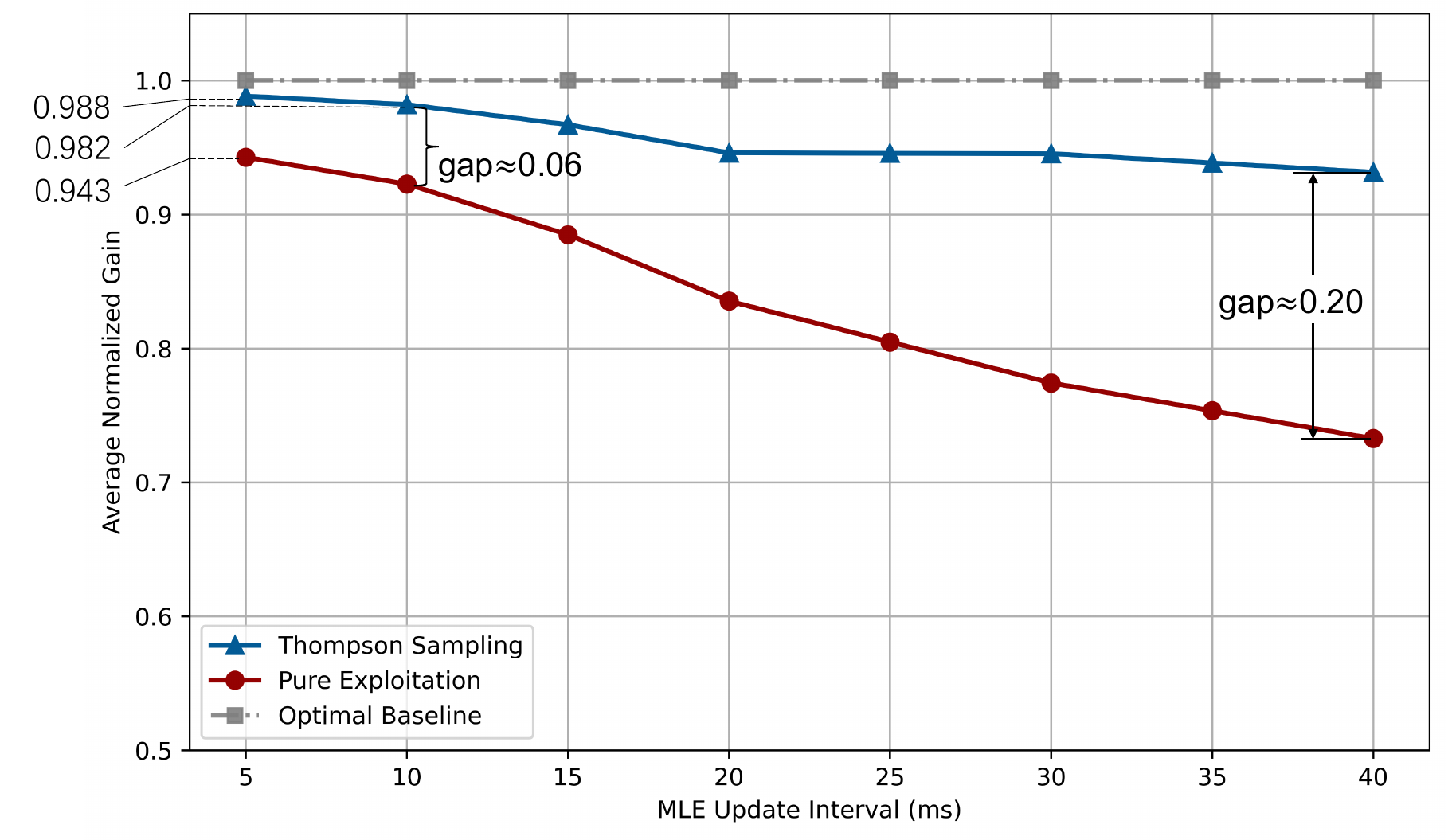}
    \caption{Tracking performance comparison of TS-based and pure exploitation schemes versus varying MLE update intervals.}
    \label{fig:normalized gain}
\end{figure}

\Cref{fig:normalized gain} compares the proposed TS-based tracking scheme with the pure exploitation scheme under different MLE update intervals $\Delta T$, using average normalized beamforming gain over $T$ as the performance metric. As $\Delta T$ increases, the tracking performance of both schemes deteriorates, as the beamformer relies on increasingly outdated motion estimates between consecutive updates. 
Nevertheless, the proposed TS-based method consistently outperforms pure exploitation over the entire trajectory for the same $\Delta T$.
When the update interval is small (e.g., $5$ ms), the TS-based scheme approaches the theoretical upper bound, whose average normalized beamforming gain is $0.988$, since the estimated trajectory parameters, $\hat{\theta}(t)$ and $\hat{r}(t)$, are frequently refined throughout the process. Conversely, a pure exploitation strategy inevitably degrades as tracking progresses due to its inability to capture new channel information through alternative probing beams. As $\Delta T$ increases, the performance gap between the two schemes becomes significantly more pronounced. Specifically, when $\Delta T = 10$ ms, the difference in average beamforming gain is about $0.06$, whereas it grows to approximately $0.20$ as $\Delta T$ increases to $40$ ms. This result underscores the necessity of uncertainty-aware exploration in near-field tracking. By sampling from the posterior distributions of $\boldsymbol{\tilde{\alpha}}_{u}$ and $\boldsymbol{\tilde{\beta}}_{u}$, the proposed method is far more resilient to the accumulation of prediction errors  compared with the pure exploration scheme. 

\begin{figure}
    \centering
    \includegraphics[width=\linewidth]{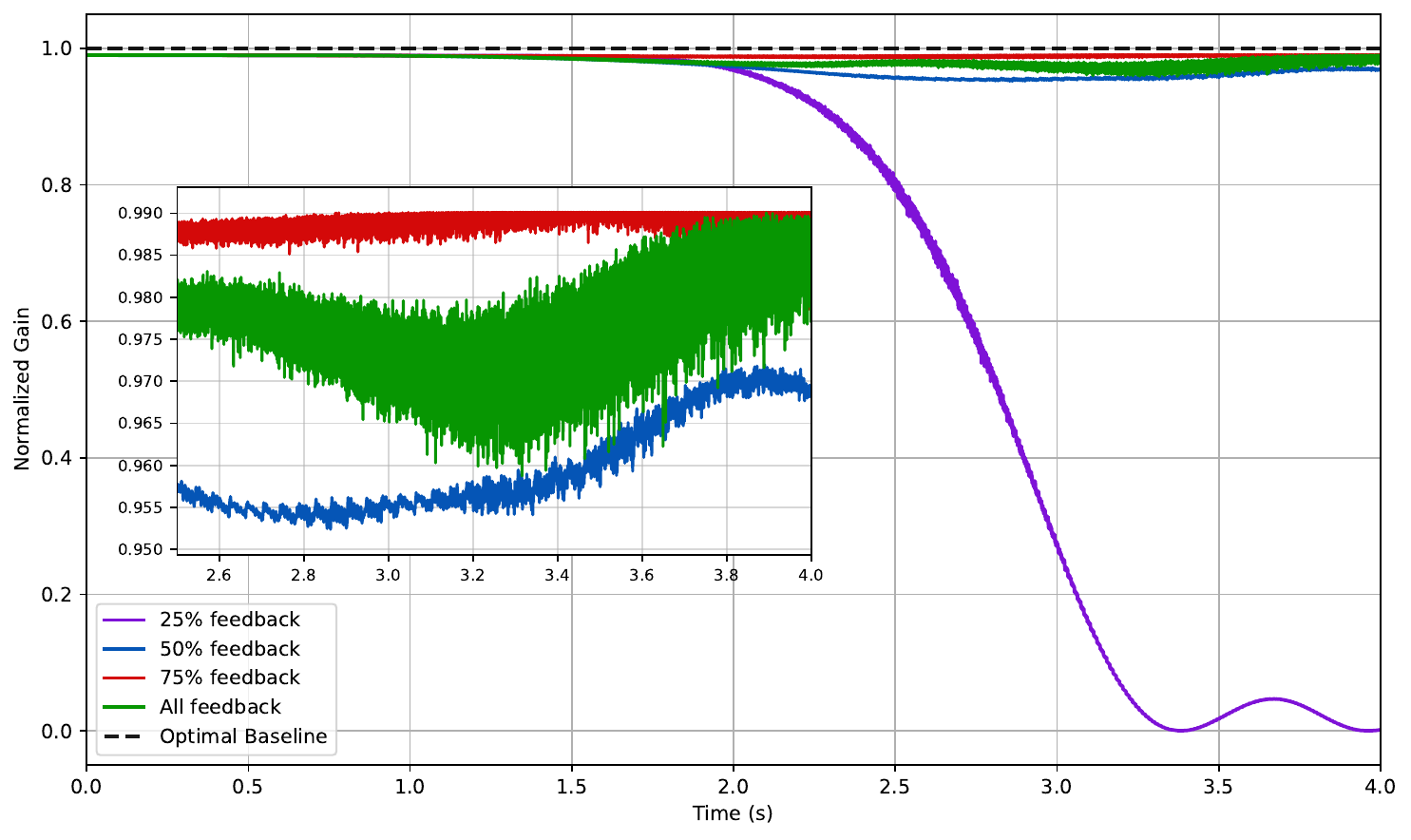}
    \caption{Tracking Performance under Different Feedback Symbol Ratio ($\Delta T=10$ms)}
    \label{fig:feedback}
\end{figure}

The proposed pilot-free method shares the closed-loop requirement for receiver feedback. Therefore, to evaluate the impact of system feedback overhead, we vary the feedback-symbol ratios ($K_F/K$) within each update interval and set $\Delta T = 10$ ms.
This interval was selected as a representative setting because, as shown in \Cref{fig:normalized gain}, the performance gap between $\Delta T = 5$ ms and $\Delta T = 10$ ms is negligible (approximately 0.006).
As illustrated in \Cref{fig:feedback}, the proposed method remains effective even when the number of feedback symbols is reduced: with only $50\%$ feedback symbols, the achieved normalized gain is almost close to that of the $100\%$-feedback strategy. 
Interestingly, the $75\%$ feedback strategy slightly outperforms the $100\%$ strategy. This occurs because the $100\%$ strategy continuously allocates resources to exploration. In contrast, the $75\%$ strategy reserves a portion of the frame for pure exploitation. Once the channel estimate is sufficiently accurate, this "greedy" allocation avoids the unnecessary perturbations inherent in TS exploration, leading to more stable beamforming. Conversely, at a $25\%$ feedback ratio, the algorithm could only track the UE for the first $2s$ due to insufficient samples for accurate estimation.

We evaluate the proposed scheme against two baselines: a classical Extended Kalman Filter (EKF)-based algorithm and the coherence-time-driven scheme in \cite{UPA_NF}. Both rely on periodic pilot-based beam sweeping. The EKF uses a near-field codebook \cite{cui2022farornf} for exhaustive searching at each pilot phase. Alternatively, \cite{UPA_NF} dynamically derives an effective beam coherence time from a tolerable gain loss and predicted user velocity. Upon coherent time expiration, the BS performs local near-field beam sweeping around the predicted position using a near field codebook subset to select the optimal codeword for subsequent transmission.

\begin{figure}[t]
    \centering
    \includegraphics[width=\linewidth]{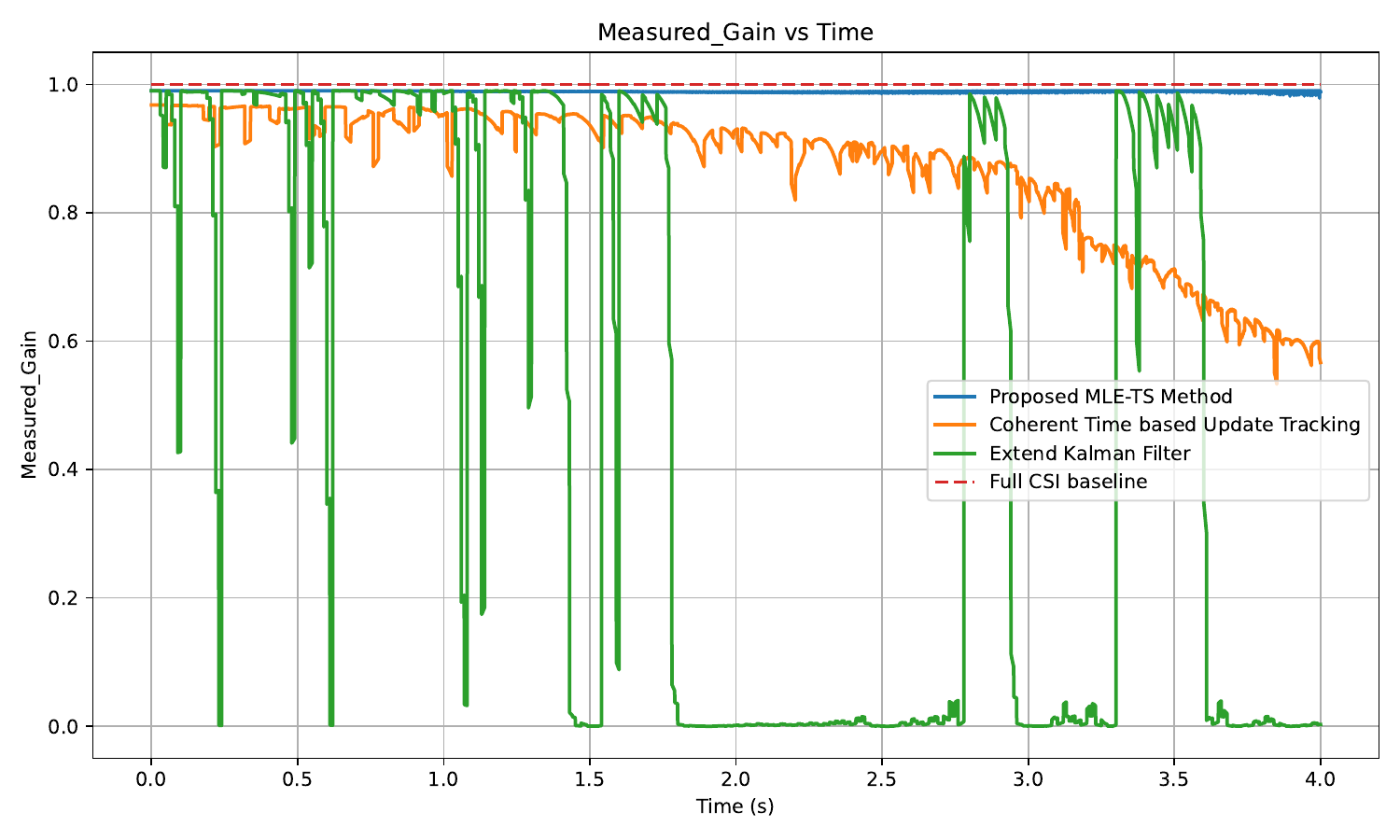}
    \caption{Normalized Gain versus Time of the proposed method and other baselines.}
    \label{fig:baselines}
\end{figure}

As illustrated in \Cref{fig:baselines}, the proposed TS-based method (with $\Delta T = 10$ ms and a 75\% feedback ratio) achieves the highest normalized beamforming gain among the baselines. It remains the most consistent with the full-CSI upper bound throughout the trajectory. Our analysis reveals the following key insights. The coherence-time-driven method generally tracks well but exhibits gradual degradation under prolonged time-varying velocities. The EKF-based scheme offers solid deterministic tracking, yet experiences sharp gain drops during pilot intervals. These drops highlight a structural drawback: the system pauses payload transmission during the exhaustive pilot phases, leading to an inevitable throughput loss. By contrast, the proposed TS method maintains a superior beamforming gain with minimal fluctuations, demonstrating a more robust and reliable tracking performance that effectively utilizes payload symbols for continuous refinement.

\section{Conclusion}\label{sec:con}
This paper presented a pilot-free framework for near-field beam tracking, integrating polynomial motion modeling, MLE, and TS. By leveraging Fisher-Information-based uncertainty quantification, the proposed method adaptively balances exploration and exploitation using payload beams, effectively eliminating the spectral efficiency loss associated with dedicated pilot-based sweeping. Numerical evaluations demonstrate that the framework achieves superior tracking robustness and significantly improved spectral efficiency compared to conventional benchmarks. These results underscore the potential of efficient solutions to address the tracking challenges of 6G near-field communication systems under user mobility.

\printbibliography

\end{document}